\documentclass{aastex631}
\usepackage[flushleft]{threeparttable}
\usepackage{natbib}
\usepackage{url}
\usepackage{placeins}
\def\UrlBreaks{\do\/\do-}

\newcommand{\mycomment}[1]{}

\begin{document}


\shortauthors{Keto \& Watters}
\shorttitle{Moving Objects in Single Satellite Images}

\title{Velocity Analysis of Moving Objects in Earth Observation Satellite Images Using Multi-Spectral Push Broom Scanning}


\author{Eric Keto}
\affiliation{Harvard University, Institute for Theory and Computation \\
60 Garden Street,
Cambridge, MA USA}

\author{Wesley Andr\'{e}s Watters}
\affiliation{Wellesley College, Department of Physics and Astronomy \\
106 Central St.,
Wellesley, MA 02481, USA}




\begin{abstract}
In this study, we present a method for detecting and analyzing the velocities of moving objects in Earth observation satellite images, specifically using data from Planet Labs' push broom scanning satellites. By exploiting the sequential acquisition of multi-spectral images, we estimate the relative differences in acquisition times between spectral bands. This allows us to determine the velocities of moving objects, such as aircraft, even without precise timestamp information from the image archive. We validate our method by comparing the velocities of aircraft observed in satellite images with those reported by onboard ADS-B transponders. The results demonstrate the potential, despite challenges posed by proprietary data limitations, of a new, useful application of commercial satellite data originally intended as an ongoing, once-daily survey of single images covering the entire land-area of the Earth. Our approach extends the applicability of satellite survey imagery for dynamic object tracking and contributes to the broader use of commercial satellite data in scientific research.

\end{abstract}

\keywords{satellite imagery, Earth observations, data analysis, image processing, push broom scanning, aircraft velocities, 
ADS-B transponders, Planet Labs, multi-spectral images.}

\section{Introduction}
In multi-spectral images made by satellites with push broom scanning, images in individual spectral bands
are acquired sequentially in time and a moving object appears in a different location in the image of each
spectral band. The velocity can be determined if the relative differences in the acquisition 
times are known for each spectral band. The effect is well known and analyzed in previous
articles. This effect has been studied by \citet{Etaya2004} and \citet{Easson2010} with images from QuickBird satellites, 
and by \citet{Heiselberg2019}
and \citet{Heiselberg2021} with images from Sentinel-2 satellites.

The satellites of Planet Labs Corporation (PLC) use push broom scanning to survey most of the land area of the
Earth with a daily revisit rate. Their archive of images, extending back several years, can 
be used to identify moving objects almost anywhere on Earth, although within a narrow time 
window of a few seconds per day. However, the acquisition times of
the images in each spectral band are not available in the PLC archive. Only an approximate time is published
for the set of images at each location. Without further assumptions this limits the determination of velocities to 
 objects that move or change with a characteristic time scale of the revisit rate of one day. \citet{Drouyer2019}
 and \citet{Chen2021} have used PLC images to identify moving vehicles and estimate traffic densities but not vehicle velocities.
 Recently, \citet{vanEtten2024} proposed a method to estimate vehicle velocities with PLC images. 
 
 In a  previous article \citep{KW23}, we proposed a method to use
 information within the images to estimate differences in the acquisition times of images
 in different spectral bands to an accuracy of
 a few tens of milliseconds. This is sufficient to study objects
moving at velocities of tens of m s$^{-1}$ rather tens of m day$^{-1}$  
with an accuracy in velocity that is limited by the positional accuracy of a few m pixel$^{-1}$ in
the PLC images.
In \citet{KW24} this procedure was followed to 
derive the altitudes of balloons from images made by PLC satellites. 
The balloons are good targets because of their
large size, high brightness, and the velocities of the balloons are low enough to approximate as zero.
In this article, we extend our analysis to trajectories of aircraft over a range of velocities of a few hundred m s$^{-1}$.
We verify our method by comparing the aircraft velocities measured from 
PLC images to the aircraft velocities reported from their onboard ADS-B transponders.  

The commercialization of space presents new opportunities in the quantity and character of available
data.  PLC satellites have unique
capabilities with higher spatial resolution ($\sim 4$ m) and a more frequent revisit rate (daily) than the 
NASA Landsat or ESA Sentinel satellites (10 - 15 m resolution and revisit rates of several days). The scientific use of
commercial data can be challenging if proprietary restrictions on information, performance and
procedures results in omissions of data such as time, one of the fundamental properties of
the universe along with space, energy and mass.

\section{Methods}

\subsection{Design of the study}

The camera on the PLC SuperDove satellites
covers eight spectral bands with a total field of view (FOV)
of $\sim 35 \times 21 \ {\rm km}^2$ and each spectral band covering one eighth the total FOV \citep{PL22}. 
With push broom scanning and a typical orbital altitude of 500 km,
the satellites observe a region on the
ground in all eight spectral bands within $\sim 3.2$ s.  We assume that the camera operates asynchronously
with the travel of the image footprint and 
with a constant rate over each image sufficient to obtain overlapping images in each spectral band. These smaller, single-band image strips,
 $\sim 35 \times 2.6 \ {\rm km}^2$,
are mosaiced by PLC into larger images with approximately the same size 
as the total FOV.  Different regions within the larger, mosaiced image then have different acquisition times, each
corresponding to the acquisition time of the smaller, overlapping strip used in each particular region.

Accounting for shifts in the acquisition times due to mosaicing is
the critical step to establishing the relative acquisition times between pixels in the mosaiced image. To test
our method of identifying the time shifts, we compare the velocities of aircraft as measured from PLC
images with the velocities reported by onboard ADS-B transponders and archived by the OpenSky Network \citep{OSN}.

In push broom scanning, the overlapping single-zone images are aligned with their long axes perpendicular to the satellite orbit. 
The effects of mosaicing on the apparent velocities of moving objects are most pronounced in tracks that
are approximately aligned with the satellite orbit. 
PLC satellites are in descending, sun-synchronous orbits with an orbital path just over 10$^\circ$ east of north.
Therefore, north-south aircraft trajectories are more likely to cross mosaicing boundaries than east-west trajectories. 
Flight paths of aircraft on close-approach or departure at the Dallas-Fort Worth International airport (DFW) are aligned with the north-south 
orientation of the major runways.
In addition DFW is one of the busiest in the world, with a high density of aircraft in the immediate airspace.
We selected the images in PLC scene ID 20220725\_161216\_34\_242b to test our method.

\subsection{Aircraft positions}

\begin{figure*}[!ht]
\begin{tabular} {ll}
\includegraphics[width=3.5in,trim={0.in 0.0in 2.5in 6.0in},clip]{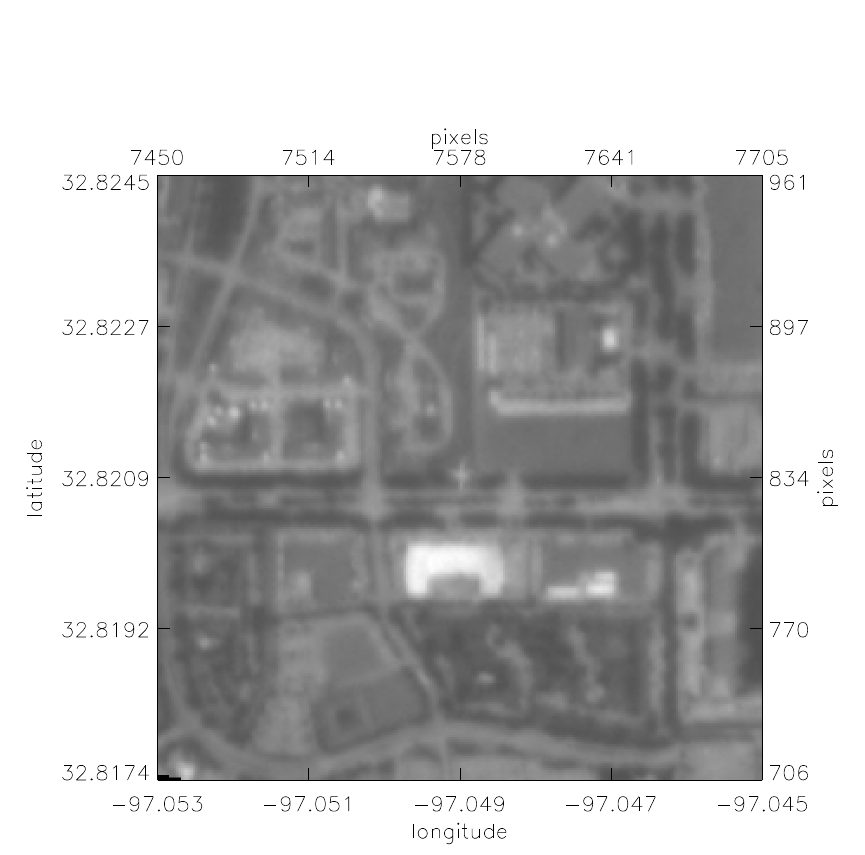} &
\includegraphics[width=3.5in,trim={0.in 0.0in 2.5in 6.0in},clip]{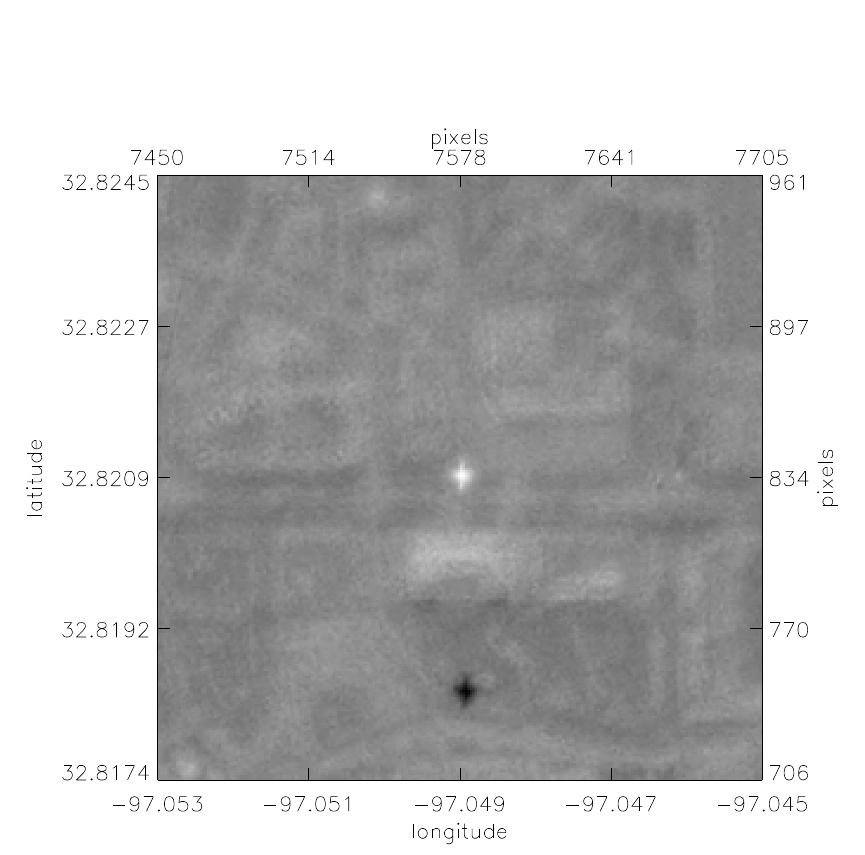} \\
\end{tabular}
\caption{Aircraft in Planet Labs Corporation (PLC) scene ID 20220725\_161216\_34\_242b. The {\it left} panel
shows the aircraft in the center of the image in a single spectral band (0 blue). The {\it right} panel
shows that the aircraft is more easily detectable in the difference of two bands (0-1). The subtraction
causes the aircraft in band 1 to appear as a negative image. The difference in position is caused
the time delay between the images in the two spectral bands. The non-moving features are suppressed
by the differencing although not entirely because the features have different brightnesses in the two spectral bands.
This particular aircraft is identified as track F in table \ref{TA}.
}
\label{locations_intro}
\end{figure*}

Moving objects can be 
isolated within a static background by differencing pairs of images acquired at different times. In the
differenced image, a moving object appears as a positive-negative pair against a suppressed
background  (figure \ref{locations_intro}). In push broom scanning,  the images acquired at different times are also in 
different spectral bands; therefore, better background suppression is achieved by differencing spectral bands
that are adjacent in wavelength.

\begin{figure*}[!ht]
\begin{tabular} {ll}
\includegraphics[width=3.5in,trim={0.in 0.0in 2.5in 6.0in},clip]{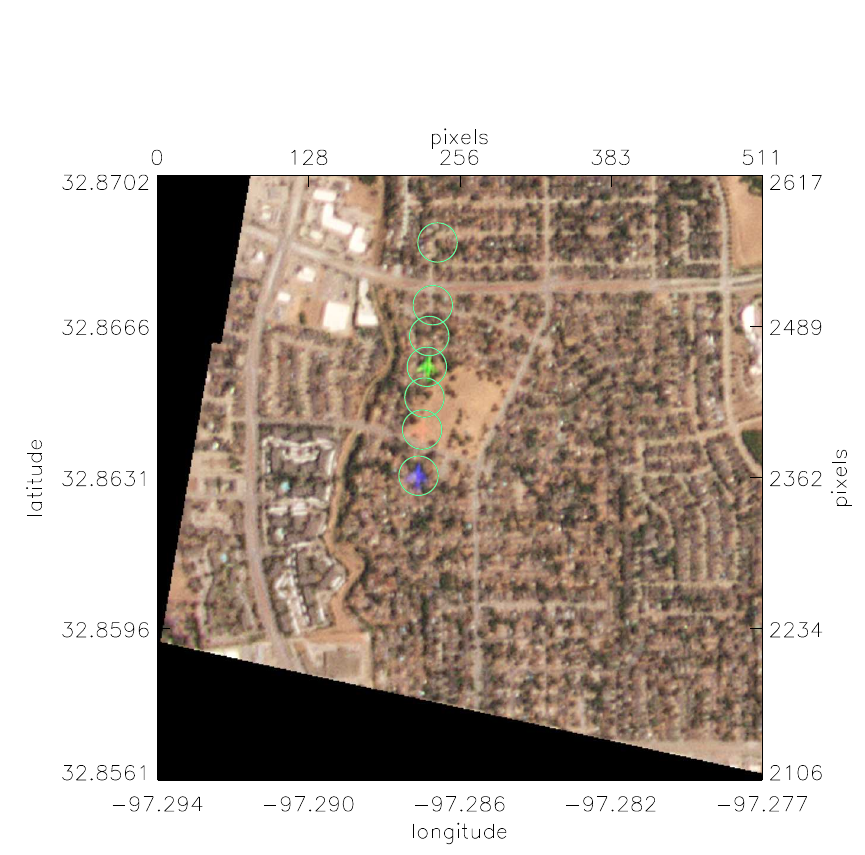} &
\includegraphics[width=3.5in,trim={0.in 0.0in 2.5in 6.0in},clip]{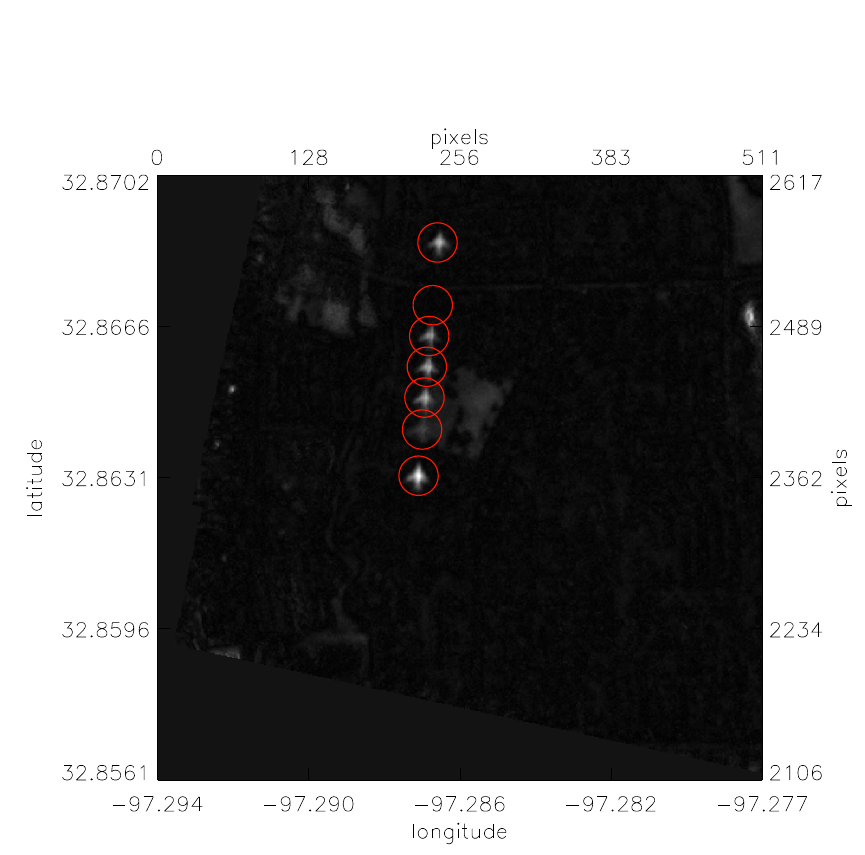} \\
\end{tabular}
\caption{Track A in table \ref{TA}.  The circles show the locations of an aircraft identified from the individual
spectral bands over a visual image ({\it left}) and over the sum of the absolute value of the seven
differenced images ({\it right}). There is no circle for the NIR band because the signal-to-noise 
is too low to allow for precise localization. The location in the NIR band is between the two
locations at the head of the track (top of the image). The spacing between the bottom two locations
is longer than the others, not counting the gap due to the missing NIR location. This longer spacing is
due to a shift in the mosaicing pattern that adds a time delay between exposures
of the camera, $\Delta t_{camera}$, to the time delay equal to the crossing time, $\Delta t_{band}$, 
of one of the single-band image strips
that are combined into a larger single-band image.
}
\label{TrackA}
\end{figure*}

\begin{figure*}[!ht]
\begin{tabular} {ll}
\includegraphics[width=3.5in,trim={0.in 0.0in 2.5in 6.0in},clip]{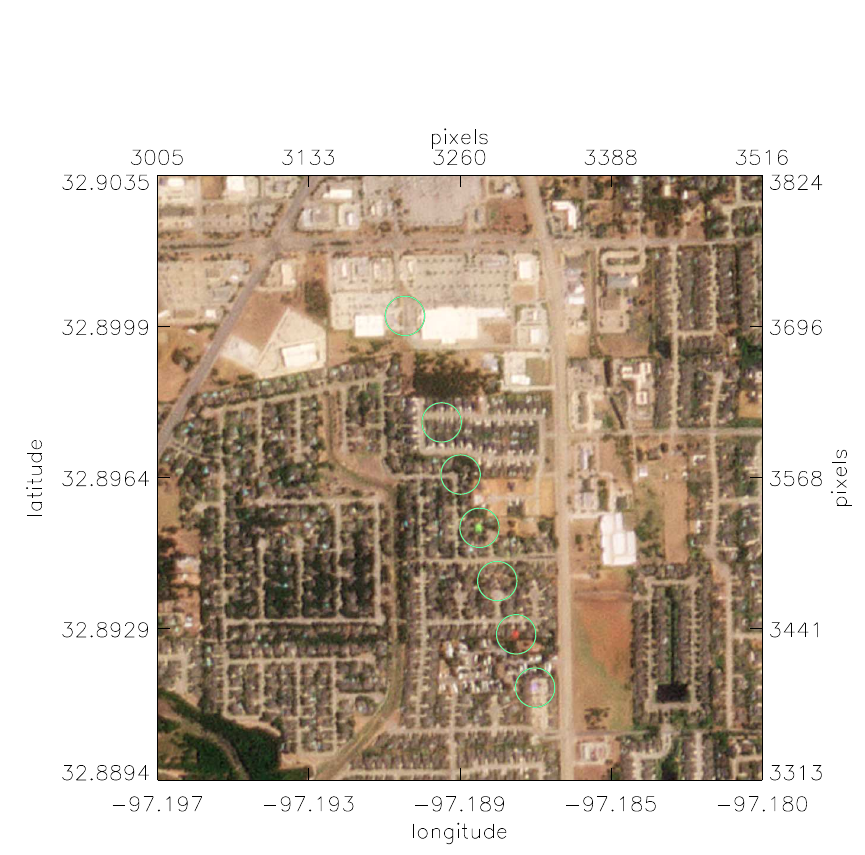} &
\includegraphics[width=3.5in,trim={0.in 0.0in 2.5in 6.0in},clip]{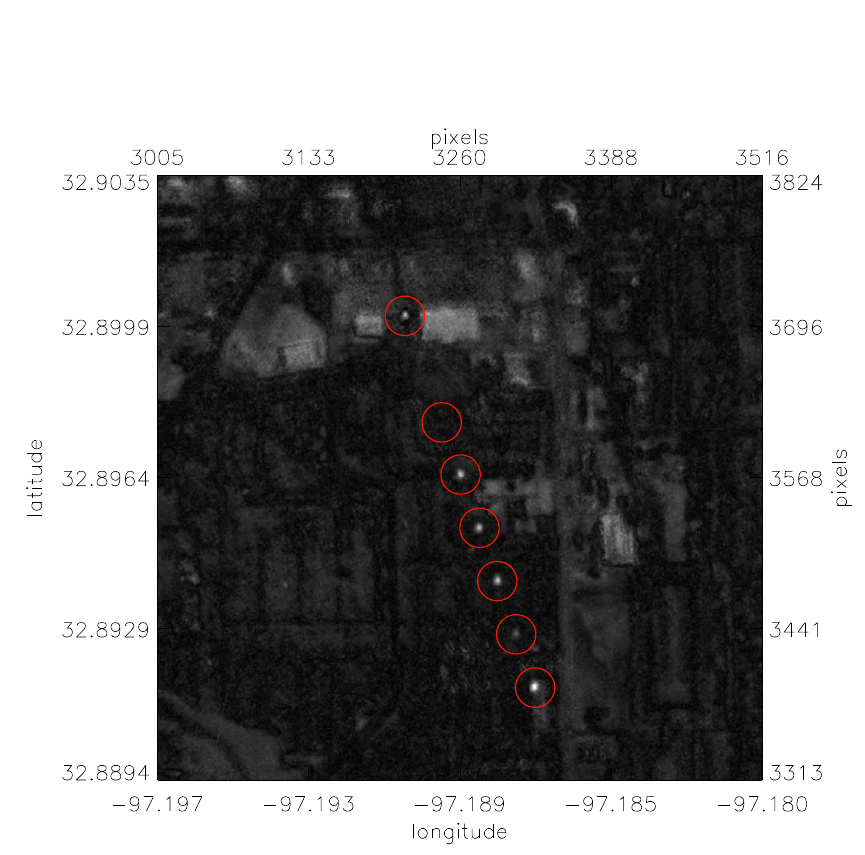} \\
\end{tabular}
\caption{Track B in table \ref{TA} in the same format as figure \ref{TrackA}. In this
case the identified locations are all equally spaced. The gap between the positions 
at the head of the track
(top of the figure) is caused by the missing NIR location. The length of the segment
with this gap is exactly twice the length of the other segments.
}
\label{TrackB}
\end{figure*}

\begin{figure*}[!ht]
\begin{tabular} {ll}
\includegraphics[width=3.5in,trim={0.in 0.0in 2.5in 6.0in},clip]{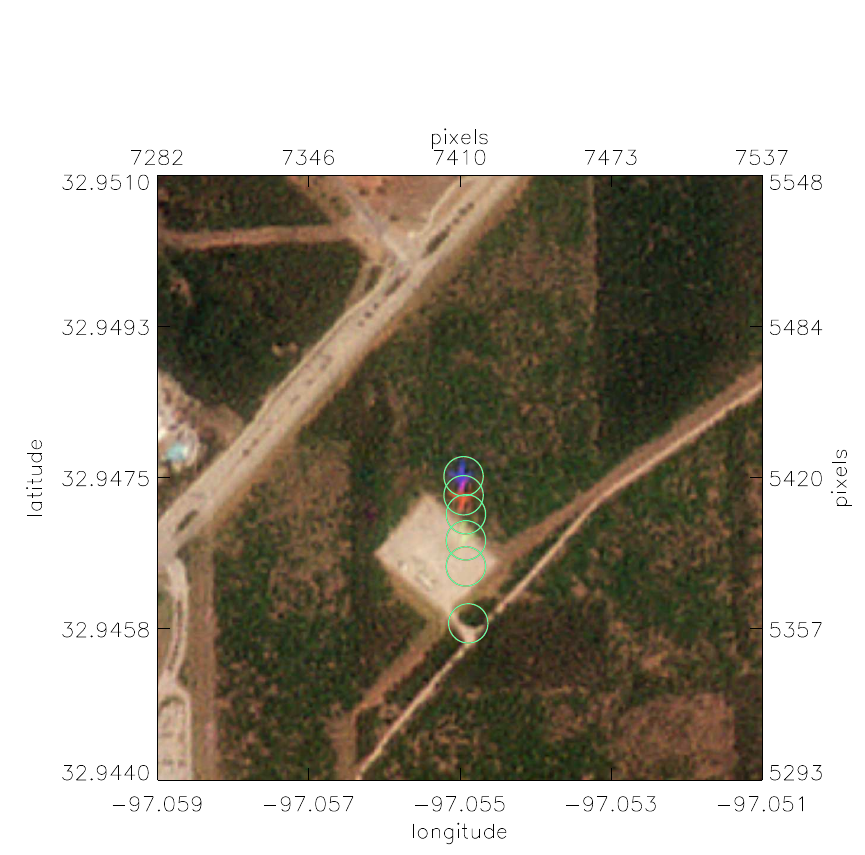} &
\includegraphics[width=3.5in,trim={0.in 0.0in 2.5in 6.0in},clip]{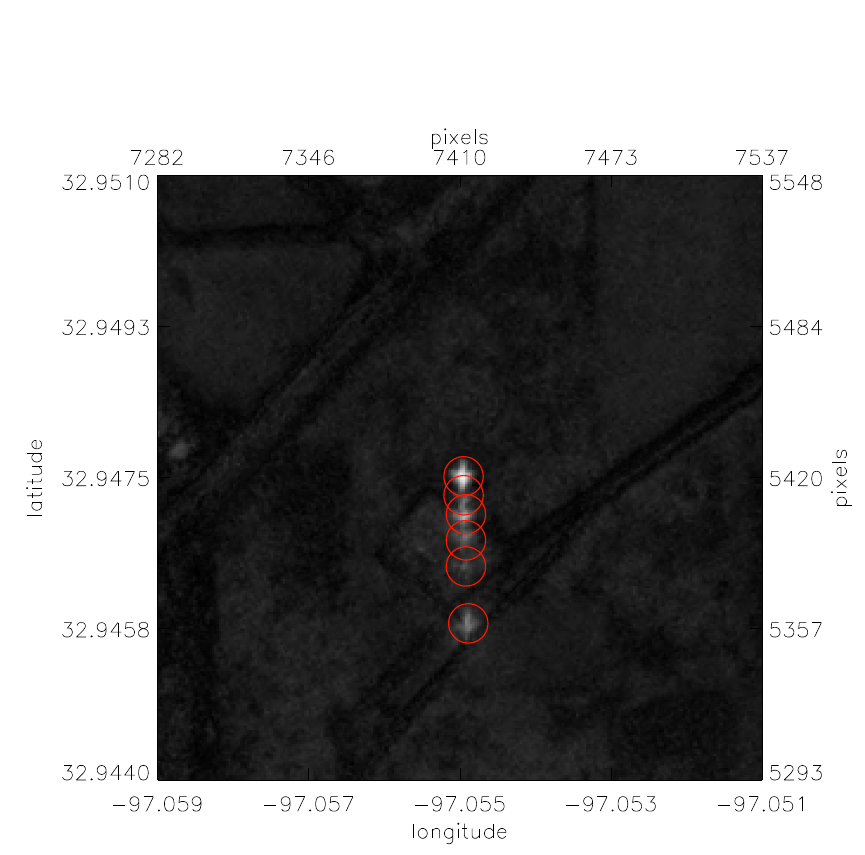} \\
\end{tabular}
\caption{Track D in table \ref{TA} in the same format as figure \ref{TrackA}. In this
case, six positions are shown because one position, in addition to the NIR, is
missing due to low signal-to-noise caused by confusion with the bright background
within the track. The NIR position would be between the two positions at the
head of the track toward the bottom of the figure. The first two segments at the
beginning of the track are shorter than the next two, each of which has an additional
time delay equal to the time between exposures of the camera.
}
\label{TrackD}
\end{figure*}

\begin{figure*}[!ht]
\begin{tabular} {ll}
\includegraphics[width=3.5in,trim={0.in 0.0in 2.5in 6.0in},clip]{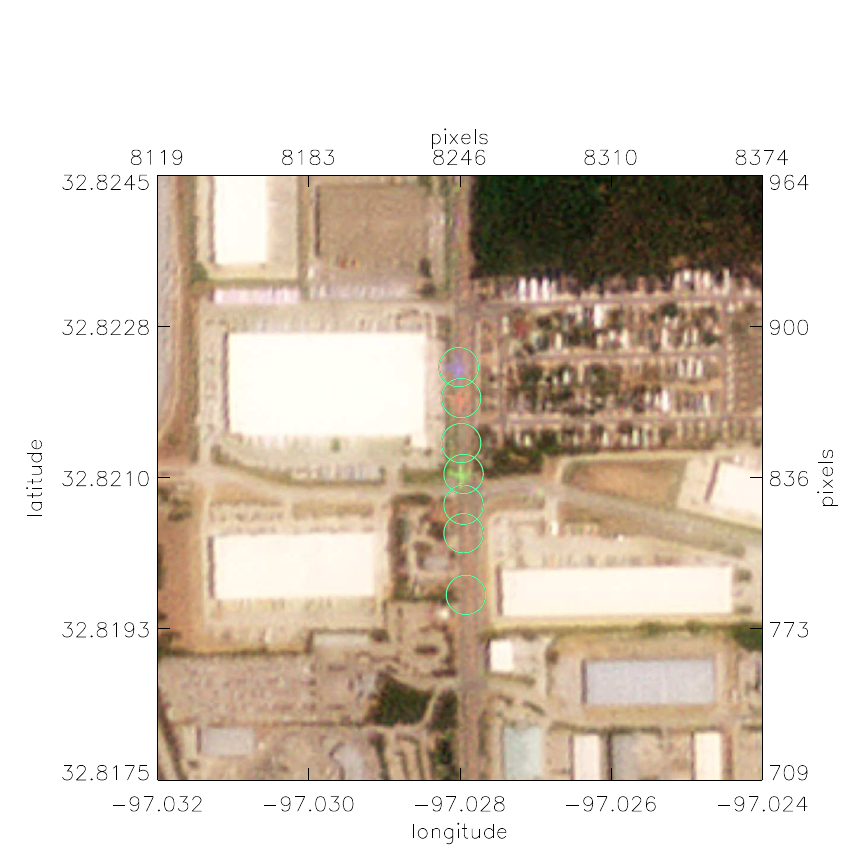} &
\includegraphics[width=3.5in,trim={0.in 0.0in 2.5in 6.0in},clip]{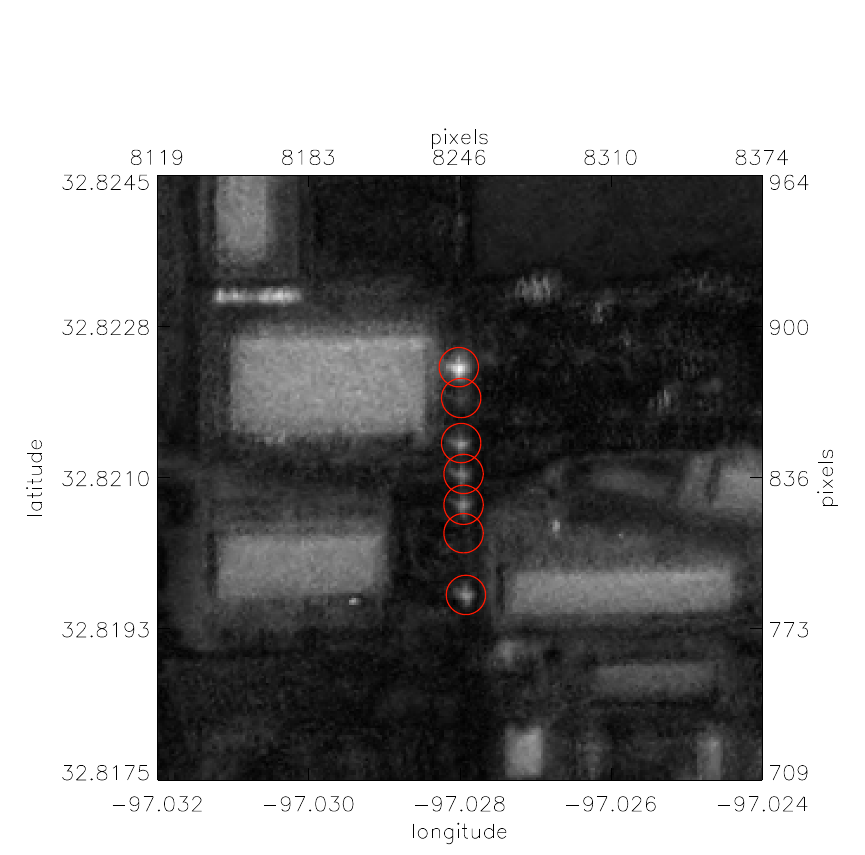} \\
\end{tabular}
\caption{Track G in table \ref{TA} in the same format as figure \ref{TrackA}. In this case
the second segment from the beginning of the track (top of image) has a longer
length and longer time delay.
}
\label{TrackG}
\end{figure*}

We determine the locations of the aircraft in the eight spectral bands by template matching. The template
is created from the average of the images of the aircraft in bands 0 and 1, both in blue visible light, because
the aircraft tend to be brighter in the shorter wavelengths. 
The positions of the aircraft are then determined from the minimum of the summed absolute difference
between the template and the seven band-differenced images (figures \ref{TrackA} - \ref{TrackG}). The aircraft
locations in each spectral band are then arranged in the order of the acquisition time that corresponds
to the spatial order of the spectral bands on the filter that is in front of the sensor. 
\begin{table}[!h]
\label{TA}
\caption{Pixel coordinates of airplanes along tracks A-H in the 8 spectral bands}
\hskip-4.0cm
\begin{tabular} {lrrrrrrrrrrrrrrrr}
\toprule
\multicolumn{1}{l}{band$^1$} 
&\multicolumn{2}{c} 1
&\multicolumn{2}{c} 5
&\multicolumn{2}{c} 2
&\multicolumn{2}{c} 3
&\multicolumn{2}{c} 4
&\multicolumn{2}{c} 6
&\multicolumn{2}{c} {7$^2$}
&\multicolumn{2}{c}  0 \\
\multicolumn{1}{l}{nm$^3$} 
&\multicolumn{2}{c}  {465-515}
&\multicolumn{2}{c}  {650-680}
&\multicolumn{2}{c}  {513-549}
&\multicolumn{2}{c}  {547-583}
&\multicolumn{2}{c}  {600-620}
&\multicolumn{2}{c}  {697-713}
&\multicolumn{2}{c}  {845-885}
&\multicolumn{2}{c}  {431-452} \\
\cline{2-3} \cline{4-5} \cline{6-7} \cline{8-9} \cline{10-11} \cline{12-13} \cline{14-15} \cline{16-17} 
ID     &  x&y  &x&y &x&y &x&y &x&y &x&y &x&y &x&y \\
\colrule
A & 92 & 2235 & 95 & 2274 & 97 & 2301 & 99 & 2327 & 101 & 2353 & 104 & 2379 & ... & ... & 108 & 2432 \\
B & 3196 & 3263 & 3180 & 3308 & 3164 & 3353 & 3149 & 3398 & 3133 & 3443 & 3117 & 3487 & ... & ... & 3086 & 3577 \\
C & 3785 & 4872 & 3786 & 4905 & 3787 & 4926 & 3788 & 4948 & 3789 & 4969 & 3790 & 4990 & ... & ... & 3792 & 5034 \\
D & 7411 & 5421 & 7411 & 5413 & 7412 & 5405 & 7412 & 5394 & 7412 & 5383 & ... & ... & ... & ... & 7413 & 5359 \\
E & 7574 & 2678 & 7574 & 2668 & 7575 & 2653 & 7576 & 2643 & 7577 & 2633 & 7579 & 2624 & ... & ... & 7579 & 2599 \\
F & 7579 & 835 & 7580 & 818 & ...      & ...    & ...      & ...    & 7580 & 781  & ...      &   ...  & ... & ... & 7581 & 744 \\
G & 8246 & 883 & 8247 & 870 & 8247 & 851 & 8248 & 838 & 8248 & 825 & 8248 & 813 & ... & ... & 8249 & 787 \\
H & 8807 & 6120 & 8808 & 6112 & 8808 & 6104 & 8808 & 6095 & 8808 & 6087 & 8809 & 6080 & ... & ... & 8809 & 6059 \\
\botrule
\end{tabular}
{\\ \footnotesize  $^1$ Spectral bands ordered by relative acquisition time.} 
{\\ \footnotesize  $^2$ Ellipses indicate that the location could not be determined because of low S/N.}
{\\ \footnotesize  $^3$ Spectral bandpass in nm.}
\end{table}

\subsection{Time delays between positions}\label{TimeDelays}

To estimate the velocities,
we require the corresponding relative acquisition times 
of the pixels at the aircraft positions in each spectral band.
 It is possible to estimate the relative acquisition times from the spacings of
the positions themselves with the assumption that the aircraft is traveling at a constant velocity. The method
is explained in  \citet{KW23,KW24}. 

The method takes advantage of necessities imposed by the push broom scanning technique 
for image acquisition.
Any pixel in the larger, mosaiced image may be derived from any one  of the partially overlapping single-band
image strips that are sequential in time. 
To allow for overlap, the time between camera exposures must be shorter than the time required for the
footprint of a strip in a single spectral band to advance across its own width. 
To minimize the quantity of data, the time between exposures should be as long as possible. For simplicity,
the time between exposures may be asynchronous with the crossing time of the single-band image strips.
Also for simplicity,  the camera may be
operating at a constant frame rate resulting in a constant time between exposures.

These requirements result in the following equation for the apparent velocity in the direction of the
satellite orbit \citep{KW23,KW24},
\begin{equation}
v_i = \frac {\Delta p_i} {\Delta t_{band} + \alpha_i\Delta  t_{camera}}
\label{eqn2}\end{equation}
where $\Delta p_i = p_i(x,y) - p_{i+1}(x,y)$ is the distance between the locations of the object in two spectral bands, $i$ and $i+1$. 
The variable $\alpha_i$ can take one of three values, $-1,0,+1$, $ \Delta  t_{camera}$ is the time between exposures, and
$\Delta t_{band}$ is the time required for the footprint of one single-band image strip to cross over another. This crossing time
depends on the orbital velocity of the satellite and the width of the footprint \citep[equation 1]{KW23}.
The set of eight positions, one for each spectral band, results in eight equations for eight unknowns, $\alpha_i$, and also
a ninth unknown, the time between exposures, $ \Delta  t_{camera}$. While the set of equations appears underdetermined, the
quantization of $\alpha_i$ represents additional information that allows a solution in two steps.
First, assuming a reasonable time between exposures just smaller than half the
crossing time of the smaller, single-band images, we solve for the eight variables, $a_i$, to minimize the deviation from constant apparent
velocity.  A good starting estimate is $ \Delta  t_{camera} = 0.184 \pm 0.007$ derived from observations of high altitude balloons  \citet{KW24}.  
Second, from the equations with non-zero $\alpha_i$, we solve for $\Delta t_{camera}$.  This value must be the same for
all the tracks in an image.

\section{Results}

Table \ref{alpha} shows the solutions for $\alpha_i$ that result from the minimization of the standard deviation
of the apparent velocities of the segments of each aircraft track given by equation \ref{eqn2}. Missing values indicate that the
the signal-to-noise of the template match is too low for a reliable determination of the position in one or
more spectral bands. In this case, the value of $\alpha_i$ applies to the segment
between valid positions. For example, four out of the eight positions of track F have good position measurements, so there are
three segments along the track. Then $\alpha_0$ applies to the segment between the positions measured in
spectral bands 1-5. The second value, $\alpha_1$ applies to the segment between the positions in spectral
bands 5-4. The last value, $\alpha_4$ applies to the segment measured between bands 4-0.

The best estimate of the time between camera exposures is $0.179 \pm 0.007$ s derived from the seven tracks, A and C-H.
Because $\alpha_i = 0$ for all segments of track B, the velocities for this track do not depend on $\Delta t_{camera}$.

\setlength{\tabcolsep}{0.5cm}
\begin{deluxetable} {lccccccc}
\tablecaption{Mosaic time correction parameters $\alpha_i$ }
\tablehead{ \colhead{Track} 	&\colhead{1-5$^a$}	&\colhead{5-2}	&\colhead{2-3}	&\colhead{3-4}	&\colhead{4-6}	&\colhead{6-0$^b$} 
} 
\startdata
A 		&1 	&0 	& 0 	& 0 	& 0 	&  0 \\
B 		& 0 	& 0 	& 0 	& 0 	& 0 	&  0 \\
C 		& 1 	& 0 	& 0 	& 0 	& 0 	&  0 \\
E 		& 0 	& 0 	& 1 	& 1 	& ... $^c$ &  0 \\
F 		& 0 	& 1 	& 0 	& 0 	& 0 	&   1 \\
G 		& 1 	& 1 	& ... 	& ... 	& 1 	&  ... \\
H 		& 0 	& 1 	& 0 	& 0 	& 0	 &  0 \\
I  		& 0 	& 0 	& 0 	& 0 	& 0 	& 1 \\
\enddata
\tablenotetext{a}{ Indicates the segment measured between the positions in spectral bands 1 and 5. The segments are arranged in time order.}
\tablenotetext{b}{ Because the NIR position (band 7) is not used, $\alpha_i$  applies to the difference in time between the positions in bands 6 and 0.}
\tablenotetext{c}{The ellipses indicate that a segment between the positions in two bands is not defined because one or both positions can not be determined due to low S/N. In this case the parameter $\alpha$ in the previous column applies to the segment from spectral band 3 to 6.}
\label{alpha}
\end{deluxetable}

Equation \ref{eqn2} and the solutions for $\alpha_i$ allow a determination of the apparent velocity for each
measured segment of the track of an aircraft. The apparent
velocity, $v_{app}$, is the sum of the true value velocity $v_{tru}$ of the aircraft and an altitude
velocity, $v_{alt}$ that is due to parallax and the motion of the satellite. The latter occurs if the object is at some altitude
above the ground and is always parallel to and in the reverse sense of the direction of the satellite orbit \citep{KW23}. The
altitude velocity depends on the altitude, velocity, and orbital angle of the satellite specified in
the parameters of a two-line element (TLE) model for the satellite orbit published daily by PLC. 
The ambiguity between the altitude and the true velocity results in a family of solutions for the true velocity
depending on altitude and apparent velocity. Table \ref{altitude-velocity} shows an example of the true velocity as a
function of altitude for track F. In this case, the altitude velocity increases with altitude because the aircraft is traveling
south, close to the opposite direction of the altitude velocity, and the true velocity therefore increases with altitude to
maintain the apparent velocity.

\begin{figure} 
\includegraphics[width=4in,trim={0.in 0.0in 2.5in 6.0in},clip] {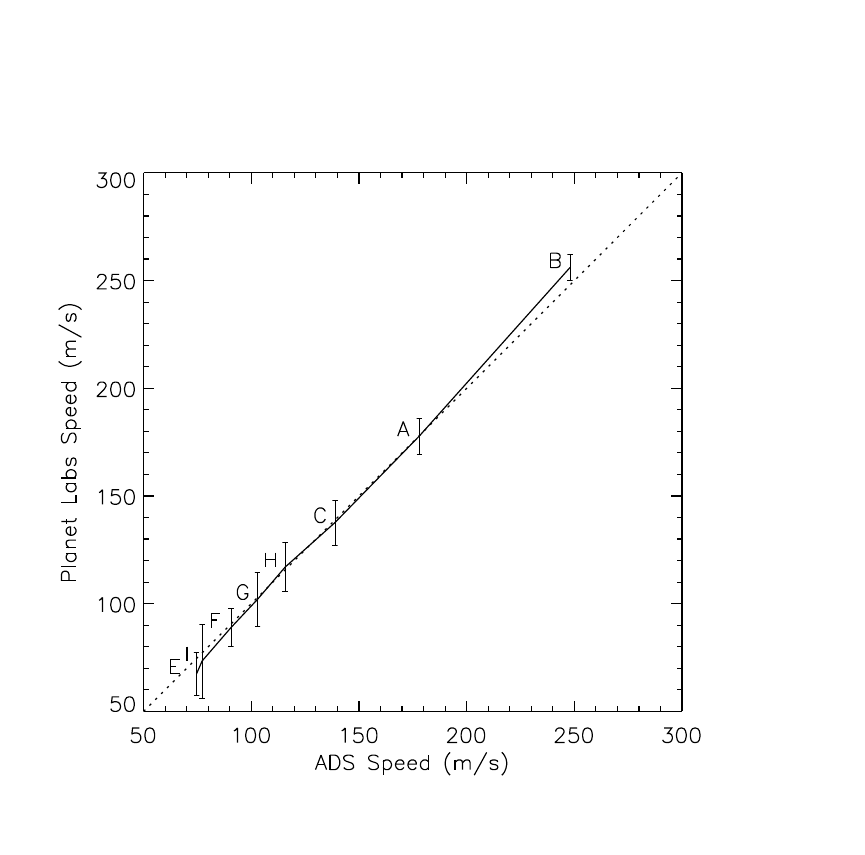} 
\caption{Comparison of velocity of aircraft as calculated from the PLC images
versus the velocity broadcast from onboard ADS-B transponders. The vertical error bars
are $3\sigma$ where $\sigma$ is the standard deviation of the velocities for each segment.
}
\label{comparison}
\end{figure}

Table \ref{comparison-velocity} and figure \ref{comparison} compare the values for the true velocities of aircraft determined from the PLC images with the 
velocities transmitted from the aircraft by their onboard ADS-B transponders. The true velocity is calculated from the apparent
velocity for the altitude indicated by the ADS-B data. 
In figure \ref{comparison}, the estimated errors for the velocities derived from the PLC images  are $3\sigma$
where $\sigma$ is the standard deviation of the velocities calculated for all segments of the track. 

The errors for the altitude (geo-altitude) and velocity from the ADS-B data are derived from onboard GPS data and assumed negligible in this calculation. 
The ADS-B data indicate a nearly constant velocity for all aircraft over a 6 s window around the 3.2 s duration required for the satellite to image all
eight bands. Although some of the aircraft are climbing or descending, 
the changes in the altitude velocity due to changes in the altitude over a 3.2 s track correspond to a negligible difference in the
calculated true velocity of less than 1 m s$^{-1}$.

The velocities derived from the PLC images depend directly on the crossing time, $\Delta t_{band}$, that depends on the
orbital velocity and the width of the footprint of one single-band image strip \citep[equation 1]{KW23}. In \citet{KW23} we assumed a
a width of 663 pixels for a single band image strip. This results in velocities from the PLC images that are systematically 3\%
lower than the ADS-B velocities indicating that the camera is using 97\% of the assumed sensor size of $8880 \times 5304$. The
velocities in table \ref{comparison-velocity} and figure \ref{comparison} are calculated with a width of 643 pixels for  a single band image strip.

\setlength{\tabcolsep}{1cm}
\begin{deluxetable}{cc}
\label{altitude-velocity}
\tablecaption{True velocity vs. altitude for track F}
\tablehead{
\colhead{Altitude$^a$} &\colhead{true velocity}	  \\
\colhead{ (m)}		&\colhead{(m s$^{-1}$)}		 
}
\startdata
          0 &   91.1 \\
   305 &   95.5 \\
   610 &   99.9 \\
   914 &  104.3 \\
  1524 &  113.1 \\
  3048 &  135.4 \\
  4572 &  158.0 \\
  6096 &  180.8 \\
  9144 &  226.8 \\
 12497 &  278.2 \\
 15240 &  320.8 \\
\enddata
\tablenotetext{a}{Above ground level (AGL) = aircraft geo-altitude - altitude of DFW. }
\end{deluxetable}

\begin{deluxetable} {ccccc}
\label{comparison-velocity}
\tablecaption{Comparison of velocities from ADS-B and PLC images}
\tablehead{
\colhead{Track} &\colhead{altitude}	&\colhead{ADS velocity}	& \colhead{PL velocity} &\colhead{std. dev.$^a$}  \\
 	&\colhead{(m)}		&\colhead{(m/s)}		&\colhead{(m/s)}		&\colhead{(m/s)} 
}
\startdata
 A  &  2985  &  178  &  177.8  &    2.8  \\
 B  & 12731  &  247  &  256.0  &    2.0  \\
 C  &  2977  &  138  &  137.7  &    3.5  \\
 E  &   158  &   74  &   67.4  &    3.3  \\
 F  &   400  &   90  &   88.9  &    3.0  \\
 G  &   760  &  102  &  102.0  &    4.2  \\
 H  &   653  &  115  &  117.0  &    3.8  \\
 I  &   371  &   77  &   73.4  &    5.7  \\
\enddata
\tablenotetext{a}{The standard deviation is a measure of the accuracy of the velocity calculated from the PLC images, specifically the differences in the velocities of the segments of the track segments. }
\end{deluxetable}

\FloatBarrier

\section{Discussion}

Our method for determining the relative acquisition times within regions of a multi-spectral image relies on an assumption of the constant velocity of
a moving object to identify time shifts introduced by PLC's mosaicing  procedure to produce the images in their archive. This assumption is not
strictly necessary and the method can be applied to objects that accelerate within the $\sim 3.2 $ s time required to observe a scene in all eight
spectral bands. This follows because the necessities imposed by push broom scanning, described in \S\ref{TimeDelays}, restrict the
variable $\alpha_i$ in equation \ref{eqn2} to discrete values of $-1, 0, +1$. 
Therefore, differences in the apparent velocities along a track that are due to acceleration  
 are generally 
incompatible with time shifts introduced by mosaicing. Accelerating objects can be recognized by an uncharacteristically
large standard deviation of the apparent velocities of the track segments. Acceleration
necessarily complicates the determination of the velocity. Our future research will explore the generalization of the method to analyze accelerating
objects and define any limitations of the method.

Conventional aircraft are not generally expected to exhibit significant acceleration within 3.2 s. Objects
with uncharacteristic flight patterns such as sudden acceleration would be of interest in our research 
with the Galileo Project  at Harvard University 
 whose goal is to collect
scientific quality data that may be useful in the search for objects of extraterrestrial origin
(https://projects.iq.harvard.edu/galileo/home).

\section{Conclusions}

Our study demonstrates a novel method for detecting and analyzing the velocities of moving objects using multi-spectral images from push broom scanning satellites.  By estimating the relative acquisition times between different spectral bands, we can accurately determine the velocities of moving objects, such as aircraft, even in the absence of precise timestamp information. Our study repurposes single-image survey data from Planet Labs Corporation SuperDove satellites for the quantitative analysis of dynamic phenomena with characteristic time scales on the order of seconds.

Key findings from our study include:
\begin{enumerate}
\item Method Validation: By comparing the velocities of aircraft derived from satellite images with those reported by onboard ADS-B transponders, we verify the accuracy and reliability of our method. The close agreement between the two sets of velocities confirms the efficacy of our approach.
\item Commercial Satellite Data Utilization: Despite the challenges posed by proprietary restrictions on commercial satellite data, our study demonstrates the potential for scientific research using these data sources. By developing techniques to infer missing temporal details, we can enhance the utility of commercial satellite imagery.
\end{enumerate}

Our work highlights the transformative impact of the commercialization of space on Earth observation. The high spatial resolution and frequent revisit rates of Planet Labs Corporation satellites offer unprecedented opportunities for detailed and timely observations of dynamic phenomena. However, the full scientific potential of these data can only be realized by addressing challenges related to data access and completeness.

\section{Acknowledgments}
Funding for access to Planet Labs data was provided by the Galileo Project at Harvard University.
We acknowledge Planet Labs Corp. for technical support.
We thank Dr. Matthew Szenher for help with the ADS-B data.

\bibliography{sat}

\end{document}